\documentclass[aps,prl,twocolumn,showpacs,superscriptaddress, preprintnumbers, hyperref]{revtex4-1}
\usepackage{latexsym}
\usepackage{amssymb}
\usepackage{graphicx}
\usepackage{amsmath}
\usepackage{bm}
\usepackage[colorlinks,
          linkcolor=black,
            citecolor=black,
            urlcolor=blue
           ]{hyperref}
\usepackage{verbatim}
\usepackage{mathrsfs}
\usepackage{extarrows}
\usepackage{comment}
\usepackage{mathtools,slashed}
\usepackage{soul}
\usepackage[toc,page]{appendix}
\usepackage[vcentermath]{youngtab}
\usepackage{multirow}
\usepackage{atbegshi,picture}
\usepackage{lipsum}
\usepackage{bbm}

\usepackage{ulem}
\usepackage{pifont}

\newcommand{\cmark}{\ding{51}}%
\newcommand{\xmark}{\ding{55}}%

\newcommand{\calH}{\mathcal{H}}

\begin{document}

\title{Lieb-Schultz-Mattis Theorem for 1D Quantum Magnets with Antiunitary Translation and Inversion Symmetries}



\author{Yuan Yao}
\email{smartyao@sjtu.edu.cn}
\affiliation{Institute of Condensed Matter Physics, School of Physics and Astronomy, Shanghai Jiao Tong University, Shanghai 200240, China}

\author{Linhao Li}
\thanks{Y.~Yao and L.~Linhao contribute equally to this work.}
\affiliation{Department of Physics and Astronomy, University of Ghent, 9000 Ghent, Belgium}
\affiliation{Institute for Solid State Physics, The University of Tokyo. Kashiwa, Chiba 277-8581, Japan}

\author{Masaki Oshikawa}
\affiliation{Institute for Solid State Physics, The University of Tokyo. Kashiwa, Chiba 277-8581, Japan}        
\affiliation{Kavli Institute for the Physics and Mathematics of the Universe (WPI),The University of Tokyo, Kashiwa, Chiba 277-8583, Japan}
\affiliation{Trans-scale Quantum Science Institute, University of Tokyo, Bunkyo-ku, Tokyo 113-0033, Japan}

\author{Chang-Tse Hsieh}         
\email{cthsieh@phys.ntu.edu.tw\\}
\affiliation{Department of Physics and Center for Theoretical Physics, National Taiwan University,
Taipei 10607, Taiwan}
\affiliation{Physics Division, National Center for Theoretical Science, National Taiwan University,
Taipei 10607, Taiwan}
\affiliation{Center for Quantum Science and Engineering, National Taiwan University, Taipei 10617, Taiwan}

\date{\today}

\begin{abstract}
We study quantum many-body systems in the presence of an exotic antiunitary translation or inversion symmetry involving time reversal.
Based on a symmetry-twisting method and spectrum robustness,
we propose that a half-integer spin chain that respects any of these two antiunitary crystalline symmetries in addition to
the discrete $\mathbb{Z}_2\times\mathbb{Z}_2$ global spin-rotation symmetry must either be gapless or possess degenerate ground states.
This explains the gaplessness of a class of chiral spin models not indicated by the Lieb-Schultz-Mattis theorem and its known extensions.
Moreover, we present symmetry classes with minimal sets of generators that give nontrivial Lieb-Schultz-Mattis-type constraints, argued by the bulk-boundary correspondence in 2D symmetry-protected topological phases as well as lattice homotopy. Our results for detecting the ingappability of 1D quantum magnets from the interplay between spin-rotation symmetries and magnetic space groups are applicable to systems with a broader class of spin interactions, including Dzyaloshinskii-Moriya and triple-product interactions.   
\end{abstract}

\maketitle

\textit{Introduction---}
Understanding and identifying various phases and the transitions among them is an essential topic in condensed matter and quantum many-body physics.
As a notable and useful concept,
the Lieb-Schultz-Mattis (LSM) theorem~\cite{Lieb:1961aa} and its generalizations~\cite{Affleck:1986aa, OYA1997,Oshikawa:2000aa, Hastings:2004ab} impose general constraints on quantum phases by the notion of ``ingappability'', i.e., the absence of a unique gapped ground state in the thermodynamic limit.
Specifically, the LSM theorem states that a system of spin 1/2's (electrons) with noninteger spin (filling) per unit cell in the presence of spin-rotation (charge-conservation) and lattice-translation symmetries must either be gapless or possess degenerate ground states. This theorem has been extended to systems with various types of on-site symmetries, e.g., discrete spin rotation symmetry~\cite{Chen-Gu-Wen_classification2010,Fuji-SymmetryProtection-PRB2016,Watanabe:2015aa,Ogata:2019aa,Ogata:2021aa,Yao:2021aa}, or even systems of generalized SU$(N)$ spins~\cite{Affleck:1986aa,Yao:2019aa}.
Moreover, the translation symmetry is found to be not necessary for the ingappability if the system respects a lattice inversion symmetry~\cite{Watanabe:2015aa,Po:2017aa,Else:2020aa,Ogata:2019aa,Ogata:2021aa,Yao:2022aa}.

Nevertheless, there exist critical systems whose ingappability is not dictated by the conventional LSM theorem and its known extensions.
One example is the 1D chiral scalar triple-product spin model:
\begin{equation}
\calH _\text{ch} = \sum_{r}(-1)^rJ\vec{S}_r\cdot(\vec{S}_{r+1}\times\vec{S}_{r+2}).
\label{eq:CTP}
\end{equation}
Such a three-spin interaction naturally appears in the third-order perturbation theory of half-filled spinful Hubbard model~\cite{Bauer:2014aa,Schmoll:2019aa}. An infinite-size density matrix renormalization group study~\cite{Schmoll:2019aa} indicates that this model exhibits critical behavior and is in the same universality class of the spin-1/2 Heisenberg chain.
However, the Hamiltonian~\eqref{eq:CTP} is invariant only under the lattice translation by two sites, while it remains the SO(3) spin-rotation symmetry, meaning that the system has an integer spin per unit cell and the conventional LSM-type theorems do not provide any constraint. Therefore, the observed gaplessness in this model appears rather mysterious.
Additional examples include critical phases in models with a staggered Dzyaloshinskii-Moriya interaction (DMI)~\cite{Oshikawa:1997aa,Zhao:2003aa, Ma:2011aa,Liu:2016aa,Qiu:2017aa},
which breaks not only the single-site translation but also any spin-rotation symmetry.
For certain cases like the spin chain with the Heisenberg exchange term plus a staggered DMI, the ingappability can be understood by a unitary transformation of the model to the uniform $XXZ$ chain~\cite{Oshikawa:1997aa}, where the conventional LSM theorem applies. However, such a unitary transformation may not exist in general and we are interested in some more robust mechanism responsible for the ingappability.

\begin{table*}[t]
\centering
\begin{tabular}{|l | c | c | c | c | c | }
\hline
 &$R^\pi_z,R^\pi_x,T_1$&$\text{TR},T_1$&$R^\pi_z,R^\pi_x,T_2$&$R^\pi_z,R^\pi_xI_1$&$~I_2~$\\
\hline
$\sum_{\alpha, r}\mathcal{J}_\alpha S^\alpha_r S^\alpha_{r+1}$&\cmark&\cmark&\cmark&\cmark&\cmark\\
\hline
$\sum_{r}\vec{D}\cdot\left(\vec{S}_r\times\vec{S}_{r+1}\right)$&\xmark&\cmark&\xmark&\cmark$^\dagger$ &\xmark\\
\hline
$\sum_{r}(-1)^r\vec{D}\cdot\left(\vec{S}_r\times\vec{S}_{r+1}\right)$&\xmark&\xmark&\xmark&\xmark&\cmark\\
\hline
$\sum_rJ\vec{S}_r\cdot(\vec{S}_{r+1}\times\vec{S}_{r+2})$&\cmark&\xmark&\xmark&\xmark&\cmark\\
\hline
$\sum_r(-1)^rJ\vec{S}_r\cdot(\vec{S}_{r+1}\times\vec{S}_{r+2})$&\xmark&\xmark&\cmark&\xmark&\cmark\\
\hline
\end{tabular}
\caption{Top row:~Symmetry classes with minimal sets of generators that sufficiently imply 1D LSM-type ingappability. Here $R^\pi_{x/z}$ is the $\pi$ rotation about the $x/z$ axis, $T_1$ is single-site translation, $I_1$ is site-centered inversion, $\text{TR}$ is time reversal, $T_2 = T_1\times\text{TR}$, and $I_2 = I_1\times\text{TR}$.
First column:~a list of typical spin interactions, including the $XYZ$ exchange term and both uniform and staggered forms of Dzyaloshinskii-Moriya and scalar triple-product interactions.  Each interaction respects at least one of these symmetry classes and is thus ingappable. Here, \cmark (\xmark) means the symmetry is (not) preserved, and \cmark$^\dagger$ means the symmetry is preserved if $\vec{D}\propto\hat{z}$.
}

\label{symmetry}
\end{table*}

In this Letter,
we extend the LSM-type ingappability to systems with integer spin per unit cell or at the inversion center by considering antiunitary lattice translation and inversion symmetries, which are important for study of magnetic materials~\cite{BFB}.
As a typical example, we consider a spin chain including a chiral triple-product interaction like Eq.~\eqref{eq:CTP}. Such a system breaks single-site translation, time-reversal, and inversion symmetries but respects certain antiunitary combinations of these symmetries. 
Along with a discrete spin-rotation symmetry, this model can be shown to be ingappable, 
derived by a simple geometric method of twisted boundary conditions~\cite{Yao:2021aa,Yao:2022aa}
based on the notion of spectrum robustness against the twisting by discrete symmetry operators~\cite{Watanabe:2018aa,Yao:2021aa}.
We further show that, motivated by the bulk-boundary correspondence in symmetry-protected topological phases, the ingappability due to the antiunitary inversion symmetry can still be guaranteed even in the absence of any spin-rotation symmetry.
We also discuss anomalies in the low-energy field theories of chiral spin models, which have a close connection to the system's ingappability on the lattice, and obtain a general constraint on the underlying universality classes.
Based on these findings, we conclude a set of symmetries that ensure the LSM-type ingappability for 1D quantum magnets in TABLE~\ref{symmetry}, which can also be understood within the framework of lattice homotopy~\cite{Po:2017aa,Else:2020aa}. 
Our results extend to a broader range of spin interactions than previously identified.


\textit{Chiral spin chains and antiunitary lattice symmetries---}
We consider a spin-$s$ chain whose Hamiltonian is the $XYZ$ model plus a staggered triple-product interaction,
\begin{align}
\label{chiral}
\mathcal{H} 
=\sum_{r,\alpha}\mathcal{J}_\alpha S^\alpha_r S^\alpha_{r+1}+\sum_{r,\alpha,\beta,\gamma}(-1)^rJ_{\alpha\beta\gamma}\epsilon_{\alpha\beta\gamma}S^\alpha_rS^\beta_{r+1}S^\gamma_{r+2},
\end{align}
where $\alpha,\beta,\gamma$ are summed over $\{x,y,z\}$ with the totally antisymmetric tensor normalized by $\epsilon_{xyz}=1$.
If $\mathcal{J}_\alpha=\mathcal{J},\,\,J_{\alpha\beta\gamma}=J$,
the model possesses the full SO$(3)$ spin-rotation symmetry.
{In this isotropic limit, the Hamiltonian is given by the sum of the standard nearest-neighbor Heisenberg exchange term and the
chiral scalar triple-product interaction as in~\eqref{eq:CTP}. The model~\eqref{eq:CTP} is thus a special case of the more general model~\eqref{chiral}.}

The Hamiltonian $\mathcal{H}$ in general lacks SO($3$) symmetry,
but it remains invariant under the discrete subgroup $\mathbb{Z}_2\times\mathbb{Z}_2$ of SO($3$) generated by
\begin{eqnarray}
R^\pi_x=\exp(i\pi\sum_rS^x_r);\,\,R^\pi_z=\exp(i\pi\sum_rS^z_r).
\end{eqnarray}
On the other hand,  $\mathcal{H}$ is invariant under \textit{two-site} lattice translation $(T_1)^2$ where $T_1:\vec{S}_r\rightarrow\vec{S}_{r+1}$,
so its unit cell consists of integer spins.
The conventional LSM theorem thus does not provide nontrivial constraint on its low-energy spectrum~\cite{Chen-Gu-Wen_classification2010,Fuji-SymmetryProtection-PRB2016,Watanabe:2015aa,Ogata:2019aa,Ogata:2021aa,Yao:2021aa}.
Nevertheless,
it respects the following antiunitary \textit{one-site} lattice translation $T_2$:
\begin{eqnarray}
T_2\vec{S}_r{T_2}^{-1}=-\vec{S}_{r+1};\,\,T_2i{T_2}^{-1}=-i.
\end{eqnarray}
It is convenient to regard $T_2$ as a product of the lattice translation and the (antiunitary) time-reversal (TR) symmetry.

This model also possesses a bond-centered inversion or reflection symmetry, $\vec{S}_{r}\rightarrow\vec{S}_{-r+1}$, with effectively two spin-$s$'s at the inversion center.
However, the known LSM-like theorems imply nothing about the system's ingappability regarding this symmetry.
Nevertheless,
we can define the following antiunitary site-centered inversion:
\begin{eqnarray}
I_2\vec{S}_r{I_2}^{-1}=-\vec{S}_{-r};\,\,I_2i{I_2}^{-1}=-i ,
\end{eqnarray}
which is a symmetry of $\mathcal{H}$.
It is a product of the unitary site-centered inversion $I_1:\vec{S}_r\rightarrow\vec{S}_{-r}$ and TR,
each of which alone is broken in $\mathcal{H}$.

\textit{Symmetry-twisting method and spectrum robustness---}
Since we consider the ingappability of bulk spectra,
the chain must be closed by a boundary condition to eliminate possible boundary modes.
The periodic boundary condition (PBC) is implicitly assumed in our original Hamiltonian~(\ref{chiral}) for the chain of length $L\in2\mathbb{Z}$ since each unit cell consists of two sites.
Instead,
let us digress for a moment and consider a Hamiltonian with the symmetry twisted boundary condition (STBC) by the symmetry operation from $\mathbb{Z}_2\times\mathbb{Z}_2$:
\begin{eqnarray}\label{stbc}
\vec{S}_{L+r}\equiv R^\pi_z\vec{S}_r(R^\pi_z)^{-1}\text{ with }r=-L,\cdots,-1,
\end{eqnarray}
where the site labeling turns out to be convenient for the inversion so that the closed ``boundary'' bond is between the sites $r=-1$ and $r=0$.
Under this boundary condition,
$\mathcal{H}$ is twisted to be
\begin{eqnarray}\label{tw}
&&\mathcal{H}_\text{tw}=\sum_{r\neq-1}\mathcal{J} S_rS_{r+1}+\sum_{r\neq-2,-1}(-1)^rJ\epsilon S_rS_{r+1}S_{r+2}\nonumber\\
&&\mathcal{J} S_{-1}\tilde{S}_{0}+(-1)^{-2}J\epsilon{S}_{-2}S_{-1}\tilde{S}_{0}+(-1)^{-1}J\epsilon{S}_{-1}\tilde{S}_{0}\tilde{S}_{1}, \nonumber
\end{eqnarray}
where we have used the abbreviations $\mathcal{J}S_rS_{r+1}\equiv\sum_\alpha\mathcal{J}_\alpha S^\alpha_rS^\alpha_{r+1}$ and $J\epsilon S_rS_{r+1}S_{r+2}\equiv\sum_{\alpha,\beta,\gamma}J_{\alpha\beta\gamma}\epsilon_{\alpha\beta\gamma}S^\alpha_rS^\beta_{r+1}S^\gamma_{r+2}$ and $\tilde{\vec{S}}_r\equiv R^\pi_z \vec{S}_r (R^\pi_z)^{-1}$ is the twisted spin operator.

\begin{figure}[t]
\centering
\includegraphics[width=8.5cm,pagebox=cropbox,clip]{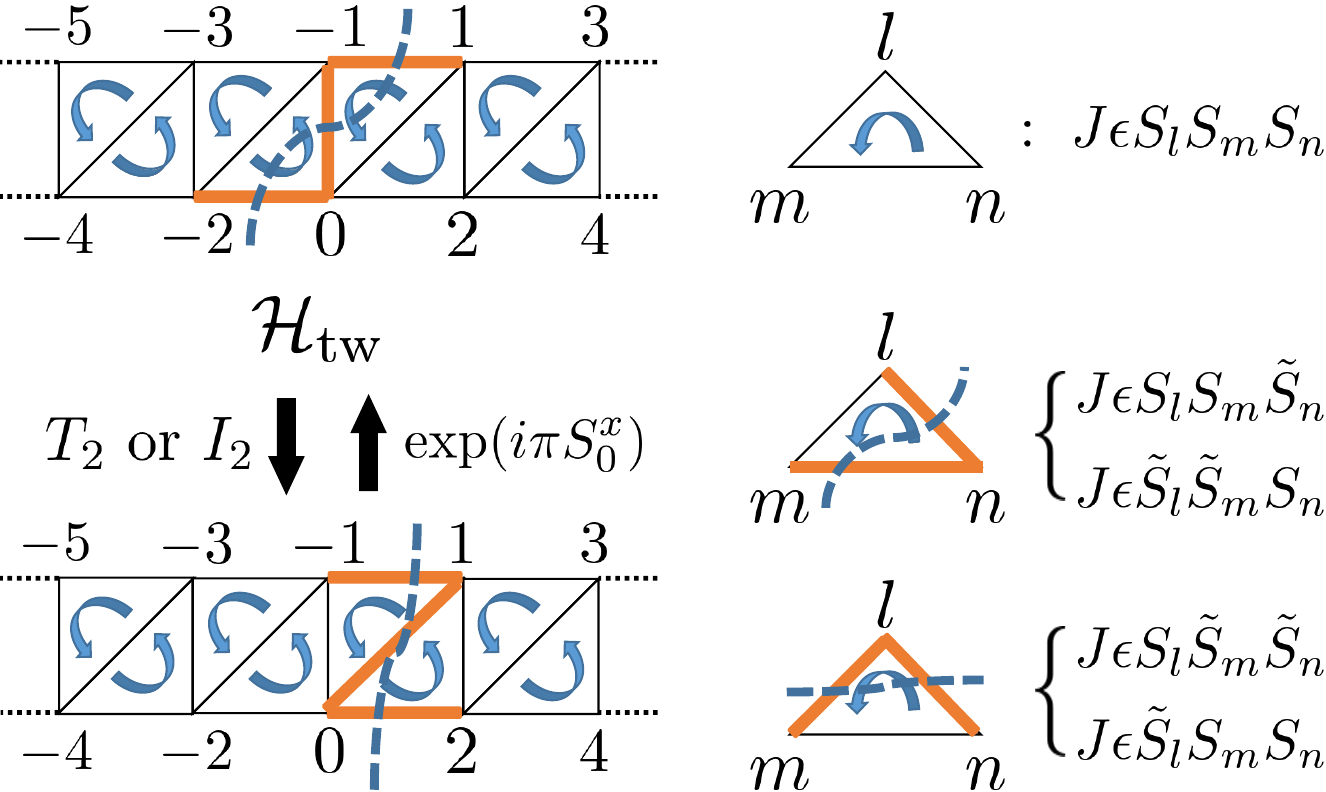}
\caption{Diagrammatic representation of the chiral spin chain~\eqref{chiral}: the sign of each triple product is indicated by the shown orientation of each triangle.
The twisted Hamiltonian $\mathcal{H}_\text{tw}$ is geometrically represented by a twisting dashed line and it possesses a modified antiunitary translation or inversion symmetry by the local transformation $\exp(i\pi S^x_0)$.
}
\label{chiral_stbc}
\end{figure}

To clarify the twisting procedure above,
we use the snakelike two-ladder diagram to represent the Hamiltonians in FIG.~\ref{chiral_stbc},
where the overall sign of each triple product can be uniquely labeled by the orientation of the triangle~\cite{Schmoll:2019aa}.
The STBC~(\ref{stbc}) can be geometrically constructed as follows.
We first draw a dashed line to separate the lattice sites $r\leq-1$ and $r\geq0$.
Then we twist all the bonds intersecting with this dashed line by acting $R^\pi_z$ on the operators on either (not both) side of the dashed line due to the $\mathbb{Z}_2$ nature of the twisting.
The twisted Hamiltonian $\mathcal{H}_\text{tw}$ explicitly breaks  $T_2$ or $I_2$,
but the breaking is soft since $\mathcal{H}_\text{tw}$ possesses the following modified symmetries by a local unitary transformation:
\begin{eqnarray}
\label{tilde}
&&\tilde{T}_2\equiv\exp(i\pi S^x_0)T_2;\,\,\tilde{I}_2\equiv\exp(i\pi S^x_0)I_2,\nonumber\\ 
&&[\tilde{T}_2,\mathcal{H}_\text{tw}]=[\tilde{I}_2,\mathcal{H}_\text{tw}]=0,
\end{eqnarray}
in addition to the original $R^\pi_{x,z}$.
Here the role played by the local transformation in Eq.~(\ref{tilde}) is to rearrange and restore the twisting configuration as illustrated in FIG.~\ref{chiral_stbc}.
We have also used the fact that the twisted bonds are moved by the $T_1$ or $I_1$ part of $T_2$ or $I_2$ whereas unaffected by their TR parts, because $R^\pi_z$ commutes with TR.
Although $\mathcal{H}_\text{tw}$ is invariant under both the modified lattice symmetries~(\ref{tilde}) and $R^\pi_z$,
they are not commutating if $s\notin\mathbb{Z}$:
\begin{eqnarray}
\label{commt}
\tilde{T}_2R^\pi_z=(-1)^{2s}R^\pi_z\tilde{T}_2;
\quad\tilde{I}_2R^\pi_z=(-1)^{2s}R^\pi_z\tilde{I}_2,
\end{eqnarray}
where the factor $(-1)^{2s}$ results from the commutator of $\exp(i\pi S^x_0)$ in Eq.~(\ref{tilde}) and $\exp(i\pi S^z_0)$ in $R^\pi_z$.
The commutator is nontrivial if $s=\mathbb{Z}+1/2$ and it implies that $\mathcal{H}_\text{tw}$ has an exact doubly degenerate spectrum for a half-integer spin chain:
for each common eigenstate $|\Psi\rangle$ of $\mathcal{H}_\text{tw}$ and $R^\pi_z$, $R^\pi_z|\Psi\rangle=\eta|\Psi\rangle$,
there exists another state $\tilde{T}_2|\Psi\rangle$ or $\tilde{I}_2|\Psi\rangle$ with the same energy but a distinct $R^\pi_z$ eigenvalue $(-1)^{2s}\eta=-\eta$, if $s\in\mathbb{Z}+1/2$.
Note that, as $(R^\pi_z)^2=1$, the eigenvalue $\eta=\pm1\in\mathbb{R}$ is unaffected by the antiunitary nature of $\tilde{T}_2$ or $\tilde{I}_2$.

We have obtained an exact degeneracy for the twisted Hamiltonian under STBC.
This implies, according to the {spectrum robustness}~\cite{Watanabe:2018aa,Yao:2021aa}, a (quasi)degeneracy of the ground states under PBC.
More precisely,
if the untwisted Hamiltonian under PBC has a unique and gapped ground state,
it can be shown that, using the quantum-transfer-matrix formulation~\cite{Betsuyaku-PRL1984},
the twisted Hamiltonian under STBC also possesses a unique gapped ground state.
{Such spectrum robustness} is also proven for U$(1)$-twisted boundary conditions in one dimension under some restriction on excited states~\cite{Watanabe:2018aa}.
Thus, 
a half-integer chiral spin chain~\eqref{chiral} under PBC must either be gapless or have a nontrivial ground-state degeneracy, rather than a unique gapped ground state that contradicts the exact degeneracy under STBC resultant from Eq.~\eqref{commt}. 

Therefore,
we arrive at the following rigorous conclusion:
\textit{the chiral spin-$s$ chain~(\ref{chiral}) under PBC must either be gapless or have a nontrivial ground-state degeneracy if $s\in\mathbb{Z}+1/2$.}

\textit{Ingappability without spin-rotation symmetries---}
The symmetry requirement for the ingappability of the chiral spin chain can be further relaxed, in the sense that
even the discrete spin-rotation symmetry can be explicitly broken.
The antiunitary inversion $I_2$ on its own is actually sufficient to guarantee the ingappability. 
In contrast, the antiunitary translation $T_2$ alone cannot ensure the ingappability.
This is because the symmetry $I_2$ is anomalous by itself---while $T_2$ is not---as our lattice model can be regarded as the boundary theory of a nontrivial 2D $I_2$-protected topological bulk consisting of paralleling Haldane chains with interchain interactions.
This bulk is nontrivial, as reasoned by the following dimensional reduction argument on crystalline symmetry-protected topological phases~\cite{Song:2017to,Huang:2017aa}:
except for the central Haldane chain positioned along the $I_2$ invariant line of the 2D system,
the rest of the bulk can be trivialized in an $I_2$ symmetric manner.
The central Haldane chain is protected by $I_2$ which reduces to the time-reversal symmetry on the $I_2$ invariant line and cannot be trivialized, thereby making the 2D bulk nontrivial.
Therefore, the chiral spin chain~\eqref{chiral}, viewed as the boundary theory of this 2D topological system, does not allow a unique gapped ground state as long as $I_2$ is respected.

Consequently, any $I_2$-symmetric half-integer spin chain must be gapless or spontaneously breaking $I_2$.
For instance, adding a staggered DMI 
$\sum_{r}(-1)^r\vec{D}\cdot(\vec{S}_r\times\vec{S}_{r+1})$ with a constant vector $\vec{D}$, which preserves the $I_2$ symmetry, to the chiral spin chain does not spoil its ingappability.
This explains the critical behaviors of the Heisenberg  or $XY$ model with a staggered DMI that have been numerically observed~\cite{Oshikawa:1997aa,Zhao:2003aa, Ma:2011aa,Liu:2016aa,Qiu:2017aa}.
Note that the conventional LSM theorem might not directly apply to these cases, as a generic $\vec{D}$ completely breaks the spin-rotation symmetry. 
It is also consistent with the proposal that the Heisenberg model with a staggered DMI can be mapped to $XXZ$ model by a unitary transformation~\cite{Oshikawa:1997aa}.
Nevertheless, the ingappability due to $I_2$ applies to a more general DMI that preserves $I_2$, e.g.
$\sum_{r}\vec{D}_{r, r+1}\cdot(\vec{S}_r\times\vec{S}_{r+1})$ 
with $\vec{D}_{r, r+1} = -\vec{D}_{-r-1, -r}$,
where the local-transformation approach is inapplicable.

The above dimensional reduction argument can also be used to show the ingappability of a half-integer spin chain with a symmetry $G$ that includes an on-site internal $\mathbb{Z}_2$ symmetry and a modified site-centered inversion symmetry $\tilde{I}$ that reduces to another internal $\mathbb{Z}_2$ symmetry at the inversion center.
For instance, $G$ can be the group generated by $R^\pi_z$ and $\tilde{I} = R^\pi_x\times I_1$. Such a symmetry is respected by, e.g.,~a uniform DMI $\sum_{r}\vec{D}\cdot(\vec{S}_r\times\vec{S}_{r+1})$ with $\vec{D}\propto\hat{z}$.

TABLE~\ref{symmetry} summarizes the results we have obtained so far, along with the known LSM-like theorems. The symmetry classes listed in this table represent minimal sets of symmetry generators that sufficiently imply the LSM-type ingappability for 1D quantum magnets of SU(2) spins.

\textit{Anomalies and ingappability---}
The LSM-type ingappability is closely related to anomalies in the low-energy effective field theories of lattice systems~\cite{Furuya:2017aa,Cheng:2016aa,Cho:2017aa,Metlitski:2018aa,Yao:2019aa,Yao:2020PRX, Cheng:2022aa}.
Let us present the above results from this viewpoint.
The chiral triple-product spin-$1/2$ chain $\mathcal{H}_\text{ch}$~\eqref{eq:CTP} is proposed to be in the universality class of SU$(2)$ level-1 conformal field theory~\cite{Schmoll:2019aa}.
The low-energy degrees of freedom can be formulated by bosonization~\cite{Affleck:1989}: $S^z_{r=x/a}\approx\frac{a}{2\pi}\partial_x\varphi-(-1)^ra_1\sin(\varphi); (S^x+iS^y)_{r=x/a}\approx\exp(i\theta)[b_0(-1)^r+b_1\sin(\varphi)],$
where $a$ is the lattice constant, $a_1,b_{0,1}$ are nonuniversal real numbers, 
and $\varphi$ and $\theta$ are compact boson fields with radius $2\pi$.
From this relation, we obtain the symmetry transformations on the boson fields~\cite{Fuji:2016aa}:
\begin{align}
R^\pi_z&:\theta\rightarrow\theta+\pi;\varphi\rightarrow\varphi;\,\,
R^\pi_x:\theta\rightarrow-\theta;\varphi\rightarrow-\varphi;\nonumber\\
\text{TR}&:\theta(x)\rightarrow\theta(x)+\pi;\varphi(x)\rightarrow-\varphi(x), \nonumber\\
T_{1(2)}&:\theta(x)\rightarrow\theta(x+a)+\pi;\varphi(x)\rightarrow(-1)\varphi(x+a)+\pi,\nonumber\\
I_1&:\theta(x)\rightarrow\theta(-x);\varphi(x)\rightarrow-\varphi(-x)+\pi. \nonumber\\
I_2&:\theta(x)\rightarrow\theta(-x)+\pi;\varphi(x)\rightarrow\varphi(-x)+\pi.
\end{align}
Here the presence of $a$ reflects the fact that the translation symmetry is not $\mathbb{Z}_2$-cyclic on the lattice.

We now discuss the anomalies in the above symmetries. First, since the TR symmetry can be realized in a $\mathbb{Z}_2\times\mathbb{Z}_2$ or SO$(3)$ invariant lattice model
{as an on-site symmetry,
e.g., in} the spin-$1/2$ Heisenberg model, there is no mixed anomaly between these two symmetries in the low-energy field theory~\cite{anomaly}.
On the other hand, $T_1$ or $I_1$ has a ($\mathbb{Z}_2$-valued) mixed anomaly with the spin-rotation symmetry $\mathbb{Z}_2\times\mathbb{Z}_2$ or SO$(3)$, due to the fact that $T_1$ or $I_1$ is broken when the $\mathbb{Z}_2\times\mathbb{Z}_2$ or SO$(3)$ global symmetry is gauged~\cite{Metlitski:2018aa,Yao:2019aa, Li:2024aa}.
This implies that $T_2=T_1\times\text{TR}$ or $I_2=I_1\times\text{TR}$ must also have a mixed anomaly with $\mathbb{Z}_2\times\mathbb{Z}_2$ or SO$(3)$~\cite{Li:2024aa}.
There are also mixed anomalies between $T_1$ and TR and between $R^\pi_z$ and $R^\pi_x\cdot I_1$, detected by the breaking of $T_1$ or $R^\pi_z$ on a crosscap state~\cite{Cho:2015aa} or a nonorientable spacetime~\cite{Sulejmanpasic:2018aa} where the orientation-reversing symmetry TR or $R^\pi_x\cdot I_1$ is gauged.
Moreover, $I_2$ has an 't Hooft anomaly by itself from the modular noninvariance of the $I_2$-orbifolded theory~\cite{Gepner:1986aa, Furuya:2017aa, EuclideanI2}.
Alternatively assuming a field-theoretical description,
we notice that $I_2$ is $\pi$-angle spacetime rotation that is \textit{orientation-preserving},
thereby associated with a \textit{unitary} $\mathbb{Z}_2$~\cite{Thorngren:2018aa}.
Its ingappability can be attributed to the nontrivial element of $H^3(\mathbb{Z}_2,\text{U}(1))=\mathbb{Z}_2$
(which signifies the role of double orientation flippings since $H^3(\mathbb{Z}_2^T,\text{U}(1))=\mathbb{Z}_1$),
compared with a conventional LSM ingappability by, e.g., $\mathbb{Z}^T_2\times\mathbb{Z}$.

By similar considerations in two dimensions,
we conjecture ingappabilities by a single antiunitary inversion with a half-integral spin at the center.
We expect it is essential in quantum-phase identifications~\cite{Yu:2024aa}.

The presence of the above anomalies is in consistency with the absence of symmetric gapping terms in the compact boson theory~\cite{Hsieh:2014ab}, indicating the LSM-type ingappability in TABLE~\ref{symmetry}~\cite{emergent}.
Furthermore, it can be shown that the even-level SU$(2)$ universality class does not possess these anomalies~\cite{Furuya:2017aa, Yao:2019aa, Sulejmanpasic:2018aa, Li:2024aa}. As a result, only odd-level universality classes can arise in the critical phases of spin-1/2 chains that preserve either $I_2$ or $T_2$ plus the (discrete) spin-rotation symmetry.

\begin{figure}[t]
\centering
\includegraphics[width=8.7cm,pagebox=cropbox,clip]{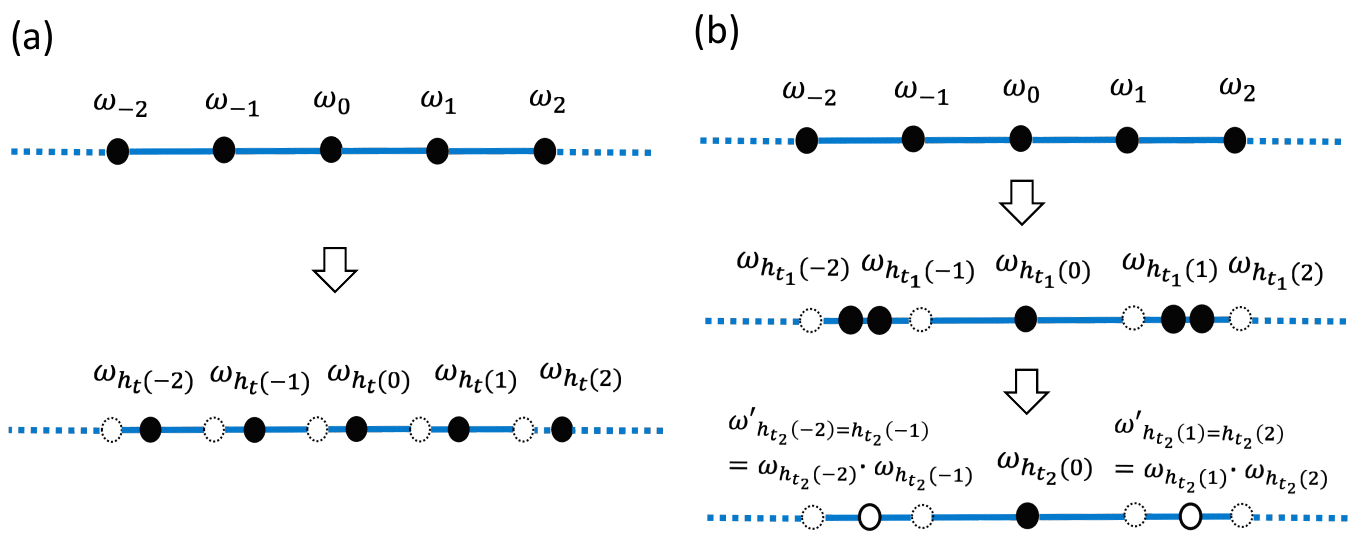}
\caption{Lattice-homotopy aspect of the LSM-type ingappability. (a) Cases involving the translation $T_1$ or $T_2$. (b) Cases involving the inversion $I_1$ or $I_2$, where site $0$ is the inversion center. Filled and non-dashed open circles denote the site spins before and after fusion, respectively.}
\label{LatticeHomotopy}
\end{figure}

\textit{Lattice homotopy for the LSM-type ingappability---}
The LSM-type ingappability for each symmetry class in TABLE~\ref{symmetry} can also be explained, in a more systematic way, within the context of lattice homotopy~\cite{Po:2017aa,Else:2020aa}. One can associate a spin system with $G$ and an {\it{anomalous texture}}, defined by an assignment of a group-cohomology element $\omega_r\in H^2(G_r, \mathrm{U}(1))$ for all lattice sites $r$.
Here, $G_r$ on each site is the subgroup of $G$ fixing site $r$, and the U(1)-coefficient group is a $G_r$-module with the antiunitary elements  (if present) of $G_r$ acting nontrivially.
A trivial anomalous texture where $\omega_r=1$ for all $r$ admits a unique gapped ground state. 
Two anomalous textures are equivalent if they are related by $G$-equivariant continuous maps $h_t(r)$ subject to $h_t(g\cdot r) = g\cdot h_t(r),~g\in G$, which represent, with the fusion condition 
$\omega'_s \equiv \Pi_{\{s_i|h_t(s_i)=s \}}~ \omega_{h_t(s_i)}$, deformations of systems under symmetric interactions.
Consequently, any system with a nontrivial anomalous texture (inequivalent to the trivial one) is ingappable.

Within lattice homotopy,  each symmetry class in column 2, 3, and 4 in TABLE~\ref{symmetry}, which involves the translation $T_1$ or $T_2$, has an anomalous texture $\{\omega_r\in H^2(G_{\text{int}}, \mathrm{U}(1))=\mathbb{Z}_2\}$, where the internal symmetry $G_{\text{int}}$ is $\mathbb{Z}_2\times\mathbb{Z}_2$ or $\mathbb{Z}_2^{\text{TR}}$. In this case, \{$\omega_{r}$\} can never be trivial for a spin chain with a half-integer spin per site, meaning that such a system is ingappable. 
On the other hand, each symmetry class in column 5 and 6 in TABLE~\ref{symmetry}, which involves the inversion $I_1$ or $I_2$, has an anomalous texture $\{\omega_c\in H^2(G_{c}, \mathrm{U}(1))=\mathbb{Z}_2~\&~\omega_{r\neq c}\in H^2(G_{r\neq c}, \mathrm{U}(1))=\mathbb{Z}_0\}$, where $G_{c}$, the $G_r$ at the inversion center(s) $r=c$, is $\mathbb{Z}_2\times\mathbb{Z}_2$ or $\mathbb{Z}_2^{\text{TR}}$ and $G_{r\neq c}$ is $\mathbb{Z}_2$ or the identity. In this case, a unique gapped ground state is prohibited for a spin chain with a half-integer spin at the inversion center(s), as \{$\omega_{r}$\} is nontrivial. FIG.~\ref{LatticeHomotopy} presents a schematic explanation of the above results.

\textit{Conclusion and discussions---}
We have presented a novel type of ingappability due to the presence of an antiunitary lattice translation or inversion symmetry. Our results, derived from the interplay between spin-rotation symmetries and magnetic space groups and summarized in TABLE~\ref{symmetry}, expand the known phases of 1D quantum magnets governed by the established LSM-like theorems to encompass a larger range of symmetries and spin interactions. 
As an illustrative example indicated by TABLE~\ref{symmetry}, a spin Hamiltonian with both uniform and staggered  scalar triple-product interactions must be ingappable, as it maintains the $I_2$ symmetry. 
Given that the staggered triple-product interaction favors the algebraic long-range antiferromagnetic order while the uniform one favors the ferromagnetic order~\cite{Schmoll:2019aa}, we conjecture a phase transition separating these two orders---an intriguing direction for future research. 
Lastly, extensions toward higher dimensional systems with more magnetic space groups involved are left as future works.

\begin{acknowledgements}
\textit{Acknowledgements---}Y.~Y. thanks the sponsorship from Yangyang Development Fund and Xiaomi Young Talents Program. L.~L. acknowledges support from Global Science Graduate Course (GSGC) program at the University of Tokyo.
The work of M. O. was supported in part by JSPS KAKENHI Grant Numbers JP19H01808, JP23H01094, and JP23K25791, and
by JST CREST Grant Number JPMJCR19T2.
C.-T. H. is supported by the Yushan (Young) Scholar Program under Grant No. NTU-111VV016 and by the National Science and Technology Council (NSTC) of Taiwan under Grant No. 112-2112-M-002-048-MY3.
\end{acknowledgements}

\bibliography{bib}

\begin{thebibliography}{45}
\bibitem [{ano()}]{anomaly}%
  \BibitemOpen
  \href@noop {} {\bibinfo  {journal} {More precisely, there is no 't Hooft
  anomaly of $(\mathbb{Z}_2\times\mathbb{Z}_2)\times\mathbb{Z}_2^{\text{TR}}$,
  i.e. no obstruction of gauging all these symmetries in the field theory}\
  }\BibitemShut {NoStop}%
\bibitem [{Euc()}]{EuclideanI2}%
  \BibitemOpen
  \href@noop {} {\bibinfo  {journal} {In two-dimensional Euclidean spacetime,
  $I_2$ is equivalent to an internal $\mathbb{Z}_2$ symmetry under a spactime
  $\pi$ rotation}\ }\BibitemShut {NoStop}%
\end{thebibliography}

\end{document}